\documentclass[10pt, twocolumn, secnumarabic, amssymb, nobibnotes, footinbib, showpacs, aps, prl, superscriptaddress]{revtex4-1}

\usepackage{graphicx}
\usepackage{dcolumn}
\usepackage{amsmath}
\usepackage{natbib}

\begin{document}


\title{State-recycling method for testing contextuality}

\author{Marek Wajs}   
\affiliation{Centre for Quantum Technologies, National University of Singapore, 3 Science Drive 2, 117543 Singapore, Singapore}

\author{Su-Yong Lee}   
\affiliation{Centre for Quantum Technologies, National University of Singapore, 3 Science Drive 2, 117543 Singapore, Singapore}

\author{Pawe\l{} Kurzy\'nski}   \email{cqtpkk@nus.edu.sg}   \affiliation{Centre for Quantum Technologies, National University of Singapore, 3 Science Drive 2, 117543 Singapore, Singapore} \affiliation{Faculty of Physics, Adam Mickiewicz University, Umultowska 85, 61-614 Pozna\'n, Poland}

\author{Dagomir Kaszlikowski}   \email{phykd@nus.edu.sg}   \affiliation{Centre for Quantum Technologies, National University of Singapore, 3 Science Drive 2, 117543 Singapore, Singapore} \affiliation{Department of Physics, National University of Singapore, 2 Science Drive 3, 117542 Singapore, Singapore}

\date{\today}


\begin{abstract}
Quantum nonlocality and contextuality are two phenomena stemming from nonclassical correlations. Whereas the former requires entanglement that is consumed in the measurement process the latter can occur for any state if one chooses a proper set of measurements. Despite this stark differences experimental tests of both phenomena were similar so far. For each run of the experiment one had to use a different copy of a physical system prepared according to the same procedure, or the system had to be brought to its initial state. Here we show that this is not necessary and that the state-independent contextuality can be manifested in a scenario in which each measurement round is done on an output state from the previous round.
\end{abstract}


\maketitle


\emph{Introduction.} In a standard Bell scenario a pair of observers share a bipartite system on which they perform local measurements \cite{Bell1964,Clauser1969}. The results of these measurements may not be explainable by local realistic theories, however to observe it one needs a quantum system prepared in an entangled state. The entanglement contained in this state is consumed during the measurement and the resulting post-measurement state is local and useless for further Bell tests. A similar effect, although more subtle, takes place in the state-dependent contextuality scenarios. For example, in the Klyachko-Can-Binicioglu-Shumovski (KCBS) test \cite{Klyachko2008} an initial state of the system can exhibit contextuality with respect to a specific set of measurements, but all the post-measurement states are noncontextual if tested in the same KCBS test. 

It is therefore natural to think of states exhibiting nonlocality or contextuality as some resourceful states, whereas the remaining states can be considered as resourceless. In this sense, the resourceful states can pass the test at the cost of becoming resourceless. Since every such test requires a sufficient amount of data to statistically determine its outcome, more than one measurement has to be performed. This requires an ensemble of resourceful states from which one draws a system in each measurement round, or a resetting procedure in which one brings back resourcefulness to the post-measurement state. 

The above interpretation makes the state-independent contextuality a different phenomenon. Every quantum state of a more than two-level system can exhibit contextuality if one prepares a special set of measurements \cite{Kochen1967,Peres1990,Mermin1990,Mermin1993,Cabello2008}. Due to this fact one cannot divide the set of all states into resourceful and resourceless since there is no resource consumption. Therefore, one is inclined to ask: \emph{How to reuse post-measurement states in some state-independent contextuality scenario?} 

There are two additional motivations behind this question. First of all, if there were an efficient method of state-recycling it would radically simplify any experimental implementations of contextuality tests in which measurements are non-destructive (e.g. trapped ion experiments \cite{Kirchmair2009}). Moreover, from the fundamental point of view it is still unclear what kind of resource contextuality is and how to quantify it with respect to some meaningful tasks \cite{Grudka2014}. Showing that post-measurement states can be efficiently reused in state-independent contextuality tests would suggest that in order to look for resourcefulness of contextuality one should not associate it to nonlocality and entanglement. A search for a new fundamental meaning of contextuality is in order.

In this work we propose a state-recycling method to investigate contextuality. We show that in the Peres-Mermin state-independent contextuality scenario \cite{Peres1990,Mermin1990,Mermin1993}, which can be also expressed in a form of an inequality \cite{Cabello2008}, the post-measurement states can be reused in the next measurement round. Moreover, because the set of post-measurement states is finite, if sufficiently many measurement rounds are performed each state will be measured in every measurement context. As a result, the measurement process on recycled states can be described as a Markov chain from which one obtains necessary statistics to evaluate the result of the test on all the states. Alternatively, it can be viewed as a test on an effective state that corresponds to the stationary distribution of the process. Eventually, we consider imperfect measurements and show that even in the presence of imperfect measurement settings our state-recycling model is still capable of detecting contextuality.


\emph{State-independent contextuality scenario.} Contextuality is a general phenomenon that can be formulated outside of quantum theory and as such it can be studied within the operational framework of black boxes \cite{Simon2001,Larsson2002}. This framework uses only the concepts of preparation, transformation and measurement. However, we want to study contextuality in realistic systems and the only known contextual systems are quantum ones. We therefore study the problem of state-recycling within the quantum formalism.

We consider the scenario commonly known as the Peres-Mermin square \cite{Peres1990,Mermin1990,Mermin1993}. It consists of nine dichotomic $\pm 1$ observables on a four-level system for which one can distinguish two two-level degrees of freedom
\begin{align}\label{MPsquare}
\begin{matrix}
A_1 = \sigma_x \otimes \openone & A_2 = \openone \otimes \sigma_y & A_3 = \sigma_x \otimes \sigma_y \\
A_4 = \openone \otimes \sigma_x & A_5 = \sigma_y \otimes \openone & A_6 = \sigma_y \otimes \sigma_x \\
A_7 = \sigma_x \otimes \sigma_x & A_8 = \sigma_y \otimes \sigma_y & A_9 = \sigma_z \otimes \sigma_z
\end{matrix}
\end{align}
where $\sigma_i$ are Pauli matrices. Each row and column consists of three jointly measurable observables. The assumption of noncontextuality states that a measurement outcome of each observable does not depend on the measurement context, i.e., whether it is measured with the observables from the same row, or with the ones from the same column. 

However, the above assumption is invalid because quantum observables are contextual. Triples of the observables in each row and column are jointly measurable because of a pairwise commutation. Note that the product $R_3$ of the three operators in the last row yields $-1$ whereas the products ($R_i$, $i=1,2$) of three operators in any other row or column ($C_i$, $i=1,2,3$) give $+1$. This gives $\prod_{ i \in \{ 1,2,3 \} } R_i C_i = -1$. However, in order to calculate this product each operator is used twice. Therefore, noncontextuality implies that no matter what value is assigned to each observable this product should be equal to $+1$.

The above \emph{paradox} can be expressed in terms of an inequality \cite{Cabello2008} 
\begin{multline}\label{Inequality}
\langle A_1 A_2  A_3 \rangle + \langle A_4 A_5 A_6 \rangle - \langle A_7 A_8 A_9 \rangle \\ + \langle A_1 A_4 A_7 \rangle + \langle A_2 A_5 A_8 \rangle + \langle A_3 A_6 A_9 \rangle \le 4
\end{multline}
The upper bound for this inequality comes from a simple optimization procedure over all possible values $A_i = \pm 1$ that takes into account the noncontextuality assumption. It is straightforward to show that for the quantum spin operators we will find the left-hand side of (\ref{Inequality}) equal to 6 because the products of all triples are $\openone$ except for $A_7 A_8 A_9 = -\openone$.


\emph{Measurements.} There are six possible measurement contexts in the Peres-Mermin scenario and in principle the measurement of each context can give one of eight possible outcomes: $+++$, $++-$, etc. However, if the measurements are perfectly implemented, there can be at most four different outcomes. This fact lies at the very root of quantum state-independent contextuality. In this case, a measurement of some triple of observables is a projection of the quantum state of the system onto one of four eigenstates of the triple, irrespective of whether the triple of measurements were done simultaneously or sequentially.  

Let us define the triple-eigenstates as $|b_{j_i}\rangle$, where $j=1,...,6$ denotes the triple and $i=1,...,4$ denotes the basis state. There are 24 different states, 16 local (corresponding to tensor products of eigenstates of $\sigma_x$ and $\sigma_y$ -- triples $j=1,\dots,4$) and 8 nonlocal (4 Bell states corresponding to the last row $j=5$ and 4 rotated Bell states corresponding to the last column $j=6$).

\begin{figure}
    \begin{center}
        \includegraphics[width=.9\columnwidth]{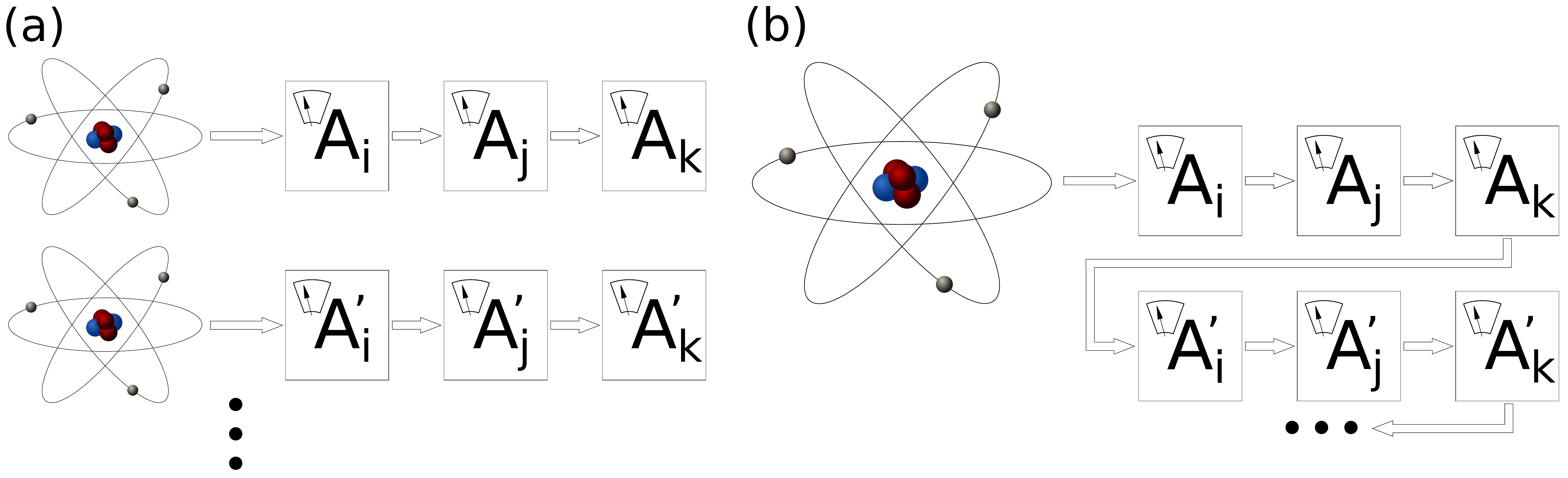} 
        \caption{Schematic representation of the standard test of contextuality in which each measurement round is performed on a different system drawn from an ensemble (a) and of the state-recycling scenario in which each measurement round is performed on the same system (b).}
        \label{fig1}
    \end{center}
\end{figure}

\emph{State-recycling scheme.} In the standard contextuality scenario one performs each measurement round on a different system drawn from an ensemble. It is assumed that the ensemble is described by the state $\rho$. For each measurement round an experimenter randomly chooses one of six measurement contexts and after many rounds one obtains enough data to evaluate the inequality (\ref{Inequality}) -- see Fig. \ref{fig1} (a). 

Here we propose a different approach in which a single copy of a system, initially prepared in some state $\rho_0$, is measured in a randomly chosen basis $j$ and as a result ends up in some post-measurement state $|b_{j_i}\rangle$ called a \emph{triple state}. Without loss of generality we can assume that $\rho_0$ is one of the 24 triple states. Then, the experimenter randomly chooses which triple $j'$ to measure next and the same system is plugged again to the measuring device. The system ends up in the triple state $|b_{j'_{i'}}\rangle$. This procedure is repeated for $N$ rounds -- see Fig. \ref{fig1} (b). 

In each round the probability for a state $|b_{j_i}\rangle$ ending up as $|b_{j'_{i'}}\rangle$ is given by
\begin{align}\label{MatrixA}
T_{j_i, j'_{i'}} = \frac{1}{6} \mathrm{Tr} \left( M_{j_i} M_{j'_{i'}} \ \right), 
\end{align}
where $M_{j_i} = |b_{j_i} \rangle \langle b_{j_i}|$. The random choice of the measurement triple gives us the factor of $1/6$. Note that the quantity $T_{j_i, j'_{i'}}$ is symmetric, i.e., $T_{j_i, j'_{i'}}=T_{j'_{i'},j_i}$ and that its evaluation requires no knowledge of any other earlier state. Therefore, the many-round measurement is a Markov chain defined on a 24-state space.  

The state of the system after $t$ measurement rounds can be represented as a 24-dimensional probability vector ${\bf p}(t)$ describing a probability distribution over all triple states. Note, that although the triple states are not all mutually orthogonal, we can distinguish the outcomes of measurements because we store information about which triple is measured and the four eigenstates of each triple are mutually orthogonal. 

The transition matrix $T$ of a process is given by Eq. (\ref{MatrixA}) -- see Appendix for an explicit form. One step of this process is given by ${\bf p}(t+1)=T{\bf p}(t)$. It is easy to verify that after many rounds the state of the system converges to the stationary distribution $\pi = T \pi$ which is a uniform distribution over all 24 triple states.


\emph{Interpretation.} We consider two alternative interpretations of the contextuality test based on the above Markov chain. In the first interpretation the state-recycling scheme leads to tests of contextuality on all 24 triple states. Note that if the process starts in a state $|b_{j_{i}}\rangle$, then after sufficiently many rounds it comes back to the same state. A crucial observation is that physically the state $|b_{j_{i}}\rangle$ that is obtained via recurrence has exactly the same properties as the initial state. This leads to a conclusion that from the experimental point of view one cannot observe a difference between state preparation in a scenario in which subsequent states  are drawn from an ensemble and a scenario in which subsequent states appear as a result of the recurrence in a state-recycling scheme.

After the recurrence the next measurement is chosen randomly, therefore for a large number of rounds one can obtain enough data to evaluate the inequality (\ref{Inequality}) for $|b_{j_{i}}\rangle$ as an input state. However, since we already know that the system state converges to the stationary distribution $\pi$, which is evenly distributed over all possible states, one can evaluate the inequality (\ref{Inequality}) for any triple state, given that sufficiently many measurements were performed.

The expected number of steps to achieve the stationary distribution $\pi$ with an accuracy $\epsilon$, i.e., the mixing time, is bounded from above by 
\begin{align}
t_{\mathrm{mix}} ( \epsilon ) \le \frac{3}{2} \log\left(\frac{24}{\epsilon}\right).
\end{align} 
See Appendix for details. For $\epsilon$ of the order of $10^{-3}$, $10^{-5}$, and $10^{-10}$ the mixing times are less than 16, 23, and 40, respectively. If we further assume that one gathers data once the system reached the distribution $\pi$ then each triple state can be obtained with equal probability $\frac{1}{24}$. One can show that an average number of sequential measurements one needs to perform on the stationary state to detect all triple states is 91 -- see Appendix.


\emph{Alternative interpretation.} Instead of considering contextuality tests on all 24 triple states, one can consider a single test of contextuality on a stationary state $\pi$. Because of indistinguishability between a stationary state and an ensemble represented by the quantum state
\begin{align}
\rho_{\pi}=\frac{1}{24}\sum_{i,j}|b_{j_i}\rangle\langle b_{j_i}| = \frac{\openone}{4},
\end{align}
measurement data collected after $\frac{3}{2} \log\left(\frac{24}{\epsilon}\right)$ steps can be considered as data measured on the state $\rho_{\pi}$. 

There is however one disadvantage in the above reasoning. One has to be aware of the fact that after the measurement the state is $|b_{j_i}\rangle$, therefore the next measurement is not performed on $\rho_{\pi}$. However, the knowledge of the triple state that is plugged into the next measuring device is conditioned on the knowledge of the previous measurement outcome. In the end, the averaging procedure that is applied to evaluate (\ref{Inequality}) effectively reduces the post-measurement state to $\rho_{\pi}$. 

Note, that in a sense this problem is also present in the standard scenarios in which an ensemble is represented by a mixed state $\sum_i p_i |\psi_i\rangle\langle \psi_i|$. The only difference is that in our case the knowledge of an exact preparation is known to the experimenter, whereas in the ensemble case this knowledge is encoded on some other inaccessible system. Therefore, the above interpretation relies on the assumption that this difference is irrelevant. In addition, in the standard ensemble scenario one can try to solve this problem by preparing a larger ensemble of entangled systems $\sum_i \sqrt{p_i} |\phi_i\rangle|\psi_i\rangle$. In this case one can use Bell-type arguments that the knowledge of the sub-ensemble preparation simply does not exist prior to a measurement on an auxiliary system. 


\emph{Errors.} While performing a sequence of measurements, we cannot ignore errors which are created by fluctuations of control parameters \cite{Szangolies2013}. These errors will lead to a detection of more than four outcomes and hence the inequality (\ref{Inequality}) is no longer maximally violated \cite{Guhne2014}. The errors occur because of a lack of control and in particular if measurement device settings fluctuate one cannot determine the post-measurement state with a perfect accuracy. 

In our error model we assume that errors come from an imperfect alignment of measuring devices. Each triple of measuring devices is perfectly aligned with a probability $p$ and completely out of control with a probability $1-p$. Therefore, the four outcomes corresponding to the triple states are detected with the probability $p$ and the remaining four outcomes, to which we refer as \emph{error states}, are detected with the probability $1-p$. 

From the point of view of the Markov chain describing the imperfect scenario, the probability space is now 48-dimensional. To calculate a new transition matrix we assume that each error state is represented by the maximally mixed state $\frac{\openone}{4}$. Note that even if in some measurement round an error state is detected, in the next round one can still detect a triple state with the probability $p$. Therefore, the state-recycling scheme is self-correcting.

The new transition matrix can be schematically represented in the form
\begin{align}\label{MatrixT}
T_{err}=
\begin{pmatrix}
\begin{array}{c|c}
  T_1 & T_2 \\
  \hline
  T_3 & T_4
\end{array}
\end{pmatrix},
\end{align}
where $T_1 = p T$ is the previous matrix (\ref{MatrixA}) multiplied by $p$. The other parts of the above matrix (\ref{MatrixT}) refer to transitions to, from and between the error states. They read $T_2 = \frac{ p }{24} \openone_{24 \times 24}$ and $T_3 = T_4 = \frac{ 1 - p }{24} \openone_{24 \times 24}$, where $\openone_{n \times n}$ is the $n \times n$ matrix with all elements equal to one. In brief, the matrix $T_2$ describes the transitions from the error states to the triple states. The opposite process is encoded by $T_3$.  Finally, $T_4$ describes transitions between the error states. Note that $T_{err}$ is stochastic as its columns sum up to 1.

The stationary distribution of $T_{err}$ corresponds to its eigenvector with the eigenvalue 1 and is $\pi = \left( \frac{p}{24}, \ldots,  \frac{p}{24}, \frac{1-p}{24}, \ldots , \frac{1-p}{24} \right)$, where the first 24 entries correspond to the triple states and the last 24 to the error states. It means that after sufficiently many steps all triple states are equally probable to be detected. This is also true for the error states. Moreover, the probability of the triple state detection is the same as the probability $p$ of the perfect setup alignment.

In the presence of noise the violation of the inequality (\ref{Inequality}) is not maximal. If we assume that the data is collected after the stationary state is obtained, the influence of noise can be easily calculated. Since the triple states appear with the probability $p$ and the error states with the probability $1-p$, one has $\langle A_i A_j A_k \rangle = 2p -1$ ($1-2p$ in the case of $\langle A_7 A_8 A_9 \rangle$). As a result, the inequality (\ref{Inequality}) reads
\begin{align}
12p - 6 \le 4.
\end{align}
It leads to the conclusion that the accuracy of the measurement setup has to be greater than $p > 5/6 \approx 0.83$ to see the violation.


\emph{Discussion.} Despite the advantage of using a single copy of a system, the state-recycling scheme can be vulnerable to a memory loophole. Apart from the loophole already discussed in Ref. \cite{Kleinmann2011}, it might be possible that the system stores information about the measurement triples measured in the past allowing for some contextual hidden variable model. This model can take advantage of this data to mimic the contextual behaviour. One can reasonably assume that in scenarios in which every measurement round is performed on a different system drawn from an ensemble such a possibility is highly unlikely since the stored information would have to be communicated from one system to the other. However, in the state-recycling scheme every measurement is performed on a single system, therefore the contextual hidden variable model would not need additional communication.

Although we do not construct any such model, we argue that its existence would require additional resources \cite{Kleinmann2011}. In the state-recycling scheme one tests a quantum system on which only two bits of classical information can be efficiently stored. On the other hand, to store information about which triple was measured one requires an additional $\log_2 6$ bits of an auxiliary memory per triple. Therefore, in order to simulate contextuality on a classical system one would require a system capable of storing more data than the original quantum system.  

One has to be aware that the problem of hidden variables in contextuality tests is slightly different from the one in Bell tests. In the case of Bell tests there exists a strong physical argument against the possibility of local realistic description (assuming the free will of observers). In order to reproduce the results of the Bell test on classical systems one would require superluminal communication, which is forbidden by the special relativity theory. On the other hand, in the case of contextuality there are no strong physical arguments against contextual hidden variables. The arguments against them are of practical nature, i.e., a simulation of contextuality on classical systems would be inefficient from the point of view of resources \cite{Kleinmann2011}. Following Occam -- it is more efficient to describe contextuality as a nonclassical effect. 

Finally, we would like to stress that our motivation is not to propose a test of contextuality that is more robust against loopholes. We propose a test whose implementation is fundamentally different and requires less experimental preparation.


\emph{Acknowledgements}. We acknowledge discussions with Adan Cabello. This work is supported by the National Research Foundation and Ministry of Education in Singapore. S.Y.L., P.K. and D.K. are also supported by the Foundational Questions Institute (FQXi).
 

\bibliography{Article.bib}


\section{Appendix}


\subsection{Transition matrix}

Here we give an explicit form of the transition matrix governing the state-recycling scheme
\begin{widetext}
\begin{align}
T=
\begin{pmatrix}
\begin{array}{cccccccccccccccccccccccc}
 \frac{1}{6} & 0 & 0 & 0 & \frac{1}{24} & \frac{1}{24} & \frac{1}{24} & \frac{1}{24} & \frac{1}{24} & \frac{1}{24} & \frac{1}{24} & \frac{1}{24} & \frac{1}{12} & \frac{1}{12} & 0 & 0 & \frac{1}{12} & 0 & \frac{1}{12} & 0 & 0 & 0 & \frac{1}{12} & \frac{1}{12} \\
 0 & \frac{1}{6} & 0 & 0 & \frac{1}{24} & \frac{1}{24} & \frac{1}{24} & \frac{1}{24} & \frac{1}{24} & \frac{1}{24} & \frac{1}{24} & \frac{1}{24} & \frac{1}{12} & \frac{1}{12} & 0 & 0 & 0 & \frac{1}{12} & 0 & \frac{1}{12} & \frac{1}{12} & \frac{1}{12} & 0 & 0 \\
 0 & 0 & \frac{1}{6} & 0 & \frac{1}{24} & \frac{1}{24} & \frac{1}{24} & \frac{1}{24} & \frac{1}{24} & \frac{1}{24} & \frac{1}{24} & \frac{1}{24} & 0 & 0 & \frac{1}{12} & \frac{1}{12} & \frac{1}{12} & 0 & \frac{1}{12} & 0 & \frac{1}{12} & \frac{1}{12} & 0 & 0 \\
 0 & 0 & 0 & \frac{1}{6} & \frac{1}{24} & \frac{1}{24} & \frac{1}{24} & \frac{1}{24} & \frac{1}{24} & \frac{1}{24} & \frac{1}{24} & \frac{1}{24} & 0 & 0 & \frac{1}{12} & \frac{1}{12} & 0 & \frac{1}{12} & 0 & \frac{1}{12} & 0 & 0 & \frac{1}{12} & \frac{1}{12} \\
 \frac{1}{24} & \frac{1}{24} & \frac{1}{24} & \frac{1}{24} & \frac{1}{6} & 0 & 0 & 0 & \frac{1}{24} & \frac{1}{24} & \frac{1}{24} & \frac{1}{24} & \frac{1}{12} & 0 & \frac{1}{12} & 0 & \frac{1}{12} & \frac{1}{12} & 0 & 0 & \frac{1}{12} & 0 & 0 & \frac{1}{12} \\
 \frac{1}{24} & \frac{1}{24} & \frac{1}{24} & \frac{1}{24} & 0 & \frac{1}{6} & 0 & 0 & \frac{1}{24} & \frac{1}{24} & \frac{1}{24} & \frac{1}{24} & 0 & \frac{1}{12} & 0 & \frac{1}{12} & \frac{1}{12} & \frac{1}{12} & 0 & 0 & 0 & \frac{1}{12} & \frac{1}{12} & 0 \\
 \frac{1}{24} & \frac{1}{24} & \frac{1}{24} & \frac{1}{24} & 0 & 0 & \frac{1}{6} & 0 & \frac{1}{24} & \frac{1}{24} & \frac{1}{24} & \frac{1}{24} & \frac{1}{12} & 0 & \frac{1}{12} & 0 & 0 & 0 & \frac{1}{12} & \frac{1}{12} & 0 & \frac{1}{12} & \frac{1}{12} & 0 \\
 \frac{1}{24} & \frac{1}{24} & \frac{1}{24} & \frac{1}{24} & 0 & 0 & 0 & \frac{1}{6} & \frac{1}{24} & \frac{1}{24} & \frac{1}{24} & \frac{1}{24} & 0 & \frac{1}{12} & 0 & \frac{1}{12} & 0 & 0 & \frac{1}{12} & \frac{1}{12} & \frac{1}{12} & 0 & 0 & \frac{1}{12} \\
 \frac{1}{24} & \frac{1}{24} & \frac{1}{24} & \frac{1}{24} & \frac{1}{24} & \frac{1}{24} & \frac{1}{24} & \frac{1}{24} & \frac{1}{6} & 0 & 0 & 0 & 0 & \frac{1}{12} & \frac{1}{12} & 0 & 0 & \frac{1}{12} & \frac{1}{12} & 0 & \frac{1}{12} & 0 & \frac{1}{12} & 0 \\
 \frac{1}{24} & \frac{1}{24} & \frac{1}{24} & \frac{1}{24} & \frac{1}{24} & \frac{1}{24} & \frac{1}{24} & \frac{1}{24} & 0 & \frac{1}{6} & 0 & 0 & \frac{1}{12} & 0 & 0 & \frac{1}{12} & 0 & \frac{1}{12} & \frac{1}{12} & 0 & 0 & \frac{1}{12} & 0 & \frac{1}{12} \\
 \frac{1}{24} & \frac{1}{24} & \frac{1}{24} & \frac{1}{24} & \frac{1}{24} & \frac{1}{24} & \frac{1}{24} & \frac{1}{24} & 0 & 0 & \frac{1}{6} & 0 & \frac{1}{12} & 0 & 0 & \frac{1}{12} & \frac{1}{12} & 0 & 0 & \frac{1}{12} & \frac{1}{12} & 0 & \frac{1}{12} & 0 \\
 \frac{1}{24} & \frac{1}{24} & \frac{1}{24} & \frac{1}{24} & \frac{1}{24} & \frac{1}{24} & \frac{1}{24} & \frac{1}{24} & 0 & 0 & 0 & \frac{1}{6} & 0 & \frac{1}{12} & \frac{1}{12} & 0 & \frac{1}{12} & 0 & 0 & \frac{1}{12} & 0 & \frac{1}{12} & 0 & \frac{1}{12} \\
 \frac{1}{12} & \frac{1}{12} & 0 & 0 & \frac{1}{12} & 0 & \frac{1}{12} & 0 & 0 & \frac{1}{12} & \frac{1}{12} & 0 & \frac{1}{6} & 0 & 0 & 0 & \frac{1}{24} & \frac{1}{24} & \frac{1}{24} & \frac{1}{24} & \frac{1}{24} & \frac{1}{24} & \frac{1}{24} & \frac{1}{24} \\
 \frac{1}{12} & \frac{1}{12} & 0 & 0 & 0 & \frac{1}{12} & 0 & \frac{1}{12} & \frac{1}{12} & 0 & 0 & \frac{1}{12} & 0 & \frac{1}{6} & 0 & 0 & \frac{1}{24} & \frac{1}{24} & \frac{1}{24} & \frac{1}{24} & \frac{1}{24} & \frac{1}{24} & \frac{1}{24} & \frac{1}{24} \\
 0 & 0 & \frac{1}{12} & \frac{1}{12} & \frac{1}{12} & 0 & \frac{1}{12} & 0 & \frac{1}{12} & 0 & 0 & \frac{1}{12} & 0 & 0 & \frac{1}{6} & 0 & \frac{1}{24} & \frac{1}{24} & \frac{1}{24} & \frac{1}{24} & \frac{1}{24} & \frac{1}{24} & \frac{1}{24} & \frac{1}{24} \\
 0 & 0 & \frac{1}{12} & \frac{1}{12} & 0 & \frac{1}{12} & 0 & \frac{1}{12} & 0 & \frac{1}{12} & \frac{1}{12} & 0 & 0 & 0 & 0 & \frac{1}{6} & \frac{1}{24} & \frac{1}{24} & \frac{1}{24} & \frac{1}{24} & \frac{1}{24} & \frac{1}{24} & \frac{1}{24} & \frac{1}{24} \\
 \frac{1}{12} & 0 & \frac{1}{12} & 0 & \frac{1}{12} & \frac{1}{12} & 0 & 0 & 0 & 0 & \frac{1}{12} & \frac{1}{12} & \frac{1}{24} & \frac{1}{24} & \frac{1}{24} & \frac{1}{24} & \frac{1}{6} & 0 & 0 & 0 & \frac{1}{24} & \frac{1}{24} & \frac{1}{24} & \frac{1}{24} \\
 0 & \frac{1}{12} & 0 & \frac{1}{12} & \frac{1}{12} & \frac{1}{12} & 0 & 0 & \frac{1}{12} & \frac{1}{12} & 0 & 0 & \frac{1}{24} & \frac{1}{24} & \frac{1}{24} & \frac{1}{24} & 0 & \frac{1}{6} & 0 & 0 & \frac{1}{24} & \frac{1}{24} & \frac{1}{24} & \frac{1}{24} \\
 \frac{1}{12} & 0 & \frac{1}{12} & 0 & 0 & 0 & \frac{1}{12} & \frac{1}{12} & \frac{1}{12} & \frac{1}{12} & 0 & 0 & \frac{1}{24} & \frac{1}{24} & \frac{1}{24} & \frac{1}{24} & 0 & 0 & \frac{1}{6} & 0 & \frac{1}{24} & \frac{1}{24} & \frac{1}{24} & \frac{1}{24} \\
 0 & \frac{1}{12} & 0 & \frac{1}{12} & 0 & 0 & \frac{1}{12} & \frac{1}{12} & 0 & 0 & \frac{1}{12} & \frac{1}{12} & \frac{1}{24} & \frac{1}{24} & \frac{1}{24} & \frac{1}{24} & 0 & 0 & 0 & \frac{1}{6} & \frac{1}{24} & \frac{1}{24} & \frac{1}{24} & \frac{1}{24} \\
 0 & \frac{1}{12} & \frac{1}{12} & 0 & \frac{1}{12} & 0 & 0 & \frac{1}{12} & \frac{1}{12} & 0 & \frac{1}{12} & 0 & \frac{1}{24} & \frac{1}{24} & \frac{1}{24} & \frac{1}{24} & \frac{1}{24} & \frac{1}{24} & \frac{1}{24} & \frac{1}{24} & \frac{1}{6} & 0 & 0 & 0 \\
 0 & \frac{1}{12} & \frac{1}{12} & 0 & 0 & \frac{1}{12} & \frac{1}{12} & 0 & 0 & \frac{1}{12} & 0 & \frac{1}{12} & \frac{1}{24} & \frac{1}{24} & \frac{1}{24} & \frac{1}{24} & \frac{1}{24} & \frac{1}{24} & \frac{1}{24} & \frac{1}{24} & 0 & \frac{1}{6} & 0 & 0 \\
 \frac{1}{12} & 0 & 0 & \frac{1}{12} & 0 & \frac{1}{12} & \frac{1}{12} & 0 & \frac{1}{12} & 0 & \frac{1}{12} & 0 & \frac{1}{24} & \frac{1}{24} & \frac{1}{24} & \frac{1}{24} & \frac{1}{24} & \frac{1}{24} & \frac{1}{24} & \frac{1}{24} & 0 & 0 & \frac{1}{6} & 0 \\
 \frac{1}{12} & 0 & 0 & \frac{1}{12} & \frac{1}{12} & 0 & 0 & \frac{1}{12} & 0 & \frac{1}{12} & 0 & \frac{1}{12} & \frac{1}{24} & \frac{1}{24} & \frac{1}{24} & \frac{1}{24} & \frac{1}{24} & \frac{1}{24} & \frac{1}{24} & \frac{1}{24} & 0 & 0 & 0 & \frac{1}{6} \\
\end{array}
\end{pmatrix}
\end{align}
\end{widetext}
It is symmetric and one can verify that the eigenvector corresponding to the eigenvalue 1, the stationary state, is of the form
\begin{align}
\pi=\frac{1}{24}(1,1,\dots,1).
\end{align}
The remaining eigenvalues are $\frac{1}{3}$ (9 eigenvalues) and $0$ (14 eigenvalues).


\subsection{Mixing time}

Next, we calculate the expected number of steps to achieve the stationary distribution $\pi$ (we follow Ref. \cite{Levin2008}). First, we need to define a distance between two probability distributions $\mu$ and $\nu$ on a finite set $\Omega$
\begin{align}
||\mu - \nu|| = \max_{A \subset \Omega} |\mu (A) - \nu (A)|.
\end{align}
In our case $\Omega = \left\{ |b_{j_i} \rangle \right\}$. Next, we define a quantity which tells us how far from stationary probability we are in the $t$-th step
\begin{align}
d(t) = \max_q || {\bf p}(t)_q - \pi_q ||,
\end{align}
where $q$ denotes the $q$-th coordinate of a probability vector. Then the mixing time, i. e., time $t$ needed for $\vec{p}(t)$ to be approximated by $\pi$, is given by
\begin{align}
t_{\mathrm{mix}} ( \epsilon ) = \min \left\{ t : d(t) \le \epsilon \right\},
\end{align}
where $\epsilon$ denotes the accuracy of the approximation.

For the above Markov chain the following inequality holds
\begin{align}\label{UpperBound}
t_{\mathrm{mix}} ( \epsilon ) \le \log \left( \frac{ 1 }{ \epsilon \pi_{\mathrm{min}} } \right) \frac{1}{1 - \lambda_{*}},
\end{align}
where $\pi_{\mathrm{min}} = \min_{q} \pi_q$ and $\lambda_{*} < 1$ is the second greatest eigenvalue of $T$. Since $\lambda_{*}=\frac{1}{3}$ we get 
\begin{align}\label{mime}
t_{\mathrm{mix}} ( \epsilon ) \le \frac{3}{2} \log \left( \frac{ 24 }{ \epsilon } \right).
\end{align}
%


\subsection{Expected number of measurements to detect all triple states} We use the solution of \emph{coupon collector's} problem to calculate the average number of times we need to keep repeating the measurement procedure to detect all $n=24$ possible triple states. Originally, the coupon collector's problem concerns how many times on average one needs to draw a coupon with replacement from some larger ensemble to collect all different coupons. There are many different generalisations of this problem, e.g. to the scenarios of non-uniformly distributed probabilities of finding each coupon or collectors not necessarily looking for all possible coupons \cite{Flajolet1992}.

Here we consider a general case of the state-recycling protocol in which errors can occur. We can formulate our problem of finding all possible triple states of the system in the state-recycling scenario as collecting coupons with almost-uniform probability distribution, meaning that an extra coupon, which does not belong to the collection (error state), might be drawn with probability $p_0$ and the probability of a coupon from collection (triple state) of size $n$ is $p_i = \frac{1 - p_0}{n} $, $i = 1, 2, \ldots, 24$. 

Thus, we treat all $24$ error states as one undesired coupon of probability $p_0 = 24 \times \frac{1-p}{24}$, where $p$ is the probability that the measurement setup is perfectly aligned. Let $T_n$ be a random variable describing a number of collected coupons until elements of every type are found for the first time. Then for $n$ coupons with almost-uniform probability distribution and one extra element the average value of $T_n$ is given by
\begin{align}
\mathbb{E} \left( T_n \right) = \frac{ n }{ 1 - p_0 } \sum_{ i = 1}^{ n } \frac{ 1 }{ i }
\end{align}
We recalled this result from \cite{Anceaume2014}, where far more general case was considered. For perfect measurements $p_0=0$ we get $T_n \approx 90.6$.


\end{document}